# Wave front adaptation using a deformable mirror for adiabatic nanofocusing along an ultrasharp gold taper

**Slawa Schmidt,[1,2] Pascal Engelke,[1,2] Björn Piglosiewicz,[1,2] Martin Esmann, [1,2] Simon F. Becker, [1,2] Kyungwan Yoo,[3] Namkyoo Park,[3] Christoph Lienau,[1,2] and Petra Groß[1,2,*]**

[1]*Institut für Physik, Carl von Ossietzky Universität, 26111 Oldenburg, Germany*
[2]*Center for Interface Science, Carl von Ossietzky Universität, 26111 Oldenburg, Germany*
[3]*Photonic Systems Laboratory, School of EECS, Seoul National University, Seoul 151-744, Korea*
[*]*Petra.Gross@uni-oldenburg.de*

**Abstract:** We describe and demonstrate the use of an adaptive wave front optimization scheme for enhancing the efficiency of adiabatic nanofocusing of surface plasmon polariton (SPP) waves along an ultrasharp conical gold taper. Adiabatic nanofocusing is an emerging and promising scheme for controlled focusing of far field light into nanometric volumes. It comprises three essential steps: SPP excitation by coupling far field light to an SPP waveguide, SPP propagation along the waveguide and adiabatic SPP nanofocusing towards a geometric singularity. For commonly used complex waveguide geometries, such as, e.g., conical metal tapers, a realistic modeling and efficiency optimization is challenging. Here, we use a deformable mirror to adaptively control the wave front of the incident far field light. We demonstrate an eight-fold enhancement in nanofocusing efficiency and analyze the shape of the resulting optimized wave front. The introduced wave front optimization scheme is of general interest for guiding and controlling light on the nanoscale.

## 1. Introduction

Adiabatic nanofocusing along conical metal tapers describes a coherent transport of optical excitations in the form of surface plasmon polariton (SPP) waves over several tens of μm and the concentration of this energy into a nanometric volume at the taper apex [1, 2]. In the adiabatic limit, i.e., if the waveguide cross section variation is slow and relative changes of the SPP wavevector are small on a scale of the SPP wavelength, radiative and reflective losses are minimized and energy transport to the apex is expected to be particularly efficient [3]. From an application point of view, adiabatic nanofocusing results in the creation of a single, dipole-like emitter, spatially localized to a few nm and with an intense optical near field. Such an emitter holds a high potential for, e.g., ultrahigh resolution optical microscopy, tip-enhanced Raman spectroscopy, or extreme ultraviolet (EUV) generation.

In the few years since its theoretical introduction, adiabatic nanofocusing has been experimentally demonstrated in different geometries, such as two-dimensional tapered waveguides [4, 5], metallic grooves, and three-dimensional conical tapers [6]. Among these, sharp conical gold tapers are of particular importance for providing nanometer-sized light sources for scattering-type scanning near field optical microscopy (s-SNOM) [6-8]. Generally, nanoslit gratings, milled onto the taper shaft at distances of up to several tens of microns from the taper apex, are used to couple far-field light onto the taper and to launch SPP waves. This distance is a compromise between achieving a virtually background-free nanolocalized light source at the apex and minimizing SPP propagation losses. With such tapers, it could so far be shown that the spatial extent of the nanofocused light source can be reduced to less than 10 nm [9, 10]. Very recently, time-resolved studies showed that the nanofocused light source could sustain pulses as short as 10 fs [9, 11, 12], which opens up all the possibilities of ultrafast micro-spectroscopy with ultrahigh temporal and spatial resolution.

While scattering and reflective losses of the SPP between grating coupling and taper apex are greatly reduced by the use of ultra-smooth, single-crystalline gold tapers [9], the over-all efficiency of adiabatic nanofocusing, defined as the ratio between the light power scattered from the tip apex and that incident onto the grating coupler, is typically below 1%, which is more than one order of magnitude below theoretical predictions [5, 13]. One reason for this restricted over-all efficiency might be a non-optimum coupling of the incident light to the fundamental mode of the SPP field on the taper shaft [14]. The focusing of an incident light pulse with a flat wave front impinging on a conical taper results in a wave front distortion of the SPP wave packet traveling down the taper shaft. Such a wave front distortion leads to partial destructive interference near the taper apex, effectively reducing the intensity of the nanofocus and thus the radiated power.

A particularly attractive possibility to correct for arbitrary wave front distortions is the use of adaptive optics, which was primarily developed for telescopes used in astronomy [15, 16]. Today adaptive optical elements are commercially available and are employed for numerous applications, e.g., to correct for thermal lensing in laser resonators [17], or for retinal imaging in ophthalmology [18].

Here we use such an adaptive optical element, namely a deformable mirror, for optimizing adiabatic nanofocusing with conical gold tapers. We demonstrate an eight-fold efficiency enhancement and show, both experimentally and theoretically, that the adapted wave front minimizes the phase mismatch between the incident light and the fundamental SPP mode of the taper.

The paper is structured as follows. In Section 2, we theoretically analyze the wave front mismatch between the incident light and the fundamental SPP mode. We determine the phase shift that is necessary to correct the mismatch. In Sections 3 and 4 we describe and demonstrate the use of a deformable mirror to manipulate the wave front of the incident pulse and to optimize this shape using an evolutionary algorithm (EA). The experimental results are presented and the resulting increase in radiated power as well as the spectral and temporal properties of the nanoconfined light source are discussed. Finally some general conclusions are given.

## 2. Theoretical consideration

### 2.1 Forward propagating light field – from the far field to the tip apex

For a first estimate, we consider a smooth, sharp conical gold taper with opening angle α, and we describe the system in spherical coordinates ($\vec{r} = (r,\theta,\varphi)$, see sketch in Fig. 1(a)). The tip apex is located at the center with $r = 0$ and the tip surface is given by coordinates with constant $\theta = \alpha$. We assume that the excitation laser light field is propagating along the *x*-direction and that the conversion of far field light to SPP waves is achieved with an idealized, infinitely narrow single slit located at $\vec{r}_{inc}(\varphi) = (r_{inc},\alpha,\varphi)$. This avoids the complexity induced by the shape of a finite size slit grating as it is usually employed in current experiments. For simplicity, we also assume that a monochromatic incident field at $\vec{r}_{inc}$ with frequency $\omega$ launches a SPP wave with a wavevector $\vec{k}_{SPP}(\varphi)$ with amplitude $k_{SPP}$, defined by the dispersion relation of a planar gold/air interface, and traveling from the slit at $\vec{r}_{inc}(\varphi) = (r_{inc}, \alpha, \varphi)$ towards the origin of the coordinate system, i.e., in opposite direction of the unit vector $\vec{e}_r$. The phase of the excited SPP field $\vec{E}_{SPP}(\vec{r}_{inc})$ is taken as identical to that of the incident laser field.

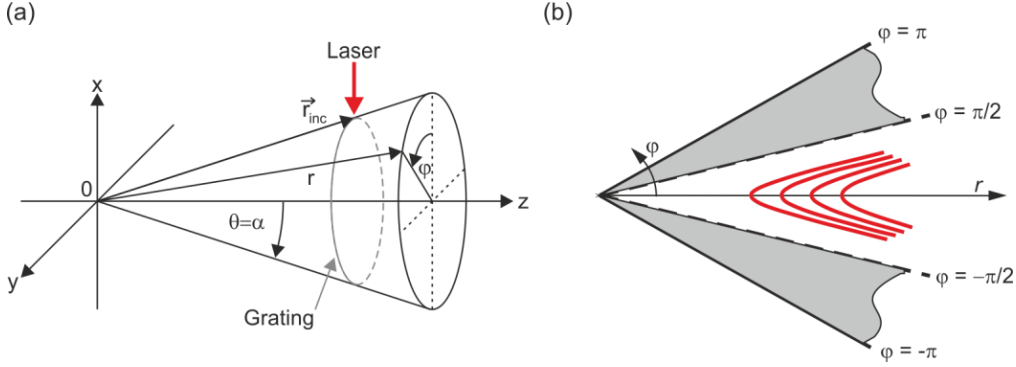

Fig.1. (a) The conical taper is described in a spherical coordinate system with the apex in the center, and the surface is defined by $\theta = \alpha$. The laser light field is incident from the top. (b) Unwrapped tip surface with a sketch of the wave fronts of the light field incident on the tip surface (red curves).

We first consider a plane wave incident field $\vec{E}_L$ with wave vector $\vec{k}_L = (-k_L, 0, 0)$ in Cartesian coordinates, where $k_L = \omega/c$ ($c$: speed of light in vacuum). The incident light travels along the negative $x$-direction as indicated by the red arrow in Fig. 1(a). Then, the wave fronts, i.e., constant phase surfaces, lie within $y$-$z$-planes, and the phase of the light field is a function of $x$:

$$\phi_L(\vec{r}) = k_L x \qquad (1)$$

The phase of the light field on the tip surface can be retrieved by performing a coordinate transform from the Cartesian to the spherical system and we obtain

$$\phi_L(\vec{r}) = k_L r \cdot \sin\alpha \cdot \cos\varphi . \qquad (2)$$

The resulting wave fronts are indicated as the red lines in the projection of the tip surface (corresponding to an unwrapped cone) in Fig. 1(b). This also gives the phase of the SPP field launched at the slit grating. SPP propagation along the taper changes the SPP phase which we can estimate as

$$\phi_{SPP}(\vec{r}) = \phi_L(\vec{r}_{inc}) + \int_{\vec{r}_{inc}}^{\vec{r}} \vec{k}_{SPP}(\varphi) d\vec{r} \qquad (3)$$

at an arbitrary point on the taper surface. It has been shown [9] that SPP dispersion has only a weak influence on the phase profile and is neglected here.

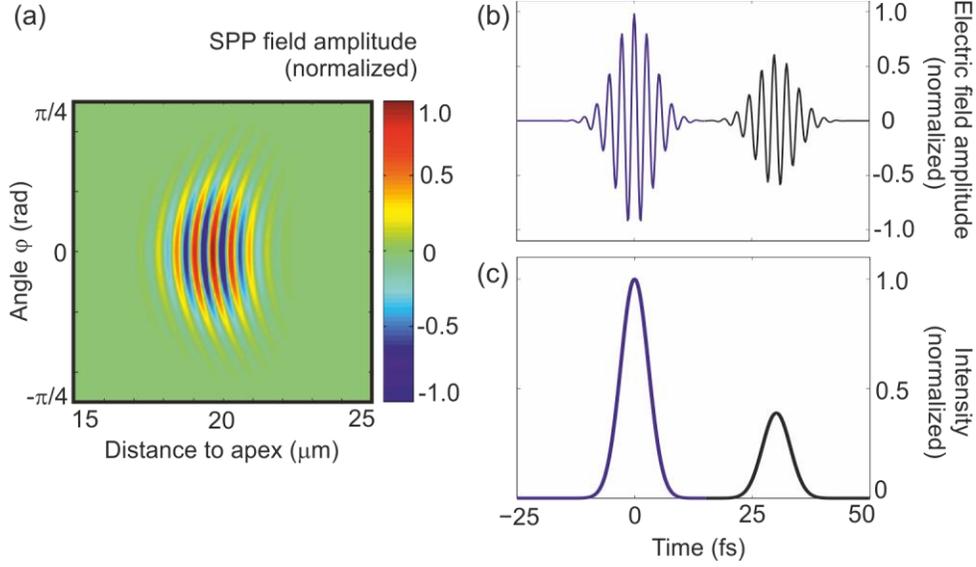

Fig. 2. Influence of the wave front distortion on the SPP electric field on the tip surface. (a) Spatial field distribution at a snap-shot in time. (b) Temporal electric field amplitude and (c) intensity profile of the distorted 7-fs SPP pulse, integrated over angle $\varphi$ (black, in the plots delayed by 30 fs) in comparison with an undistorted SPP pulse, showing the reduction in peak intensity.

It is evident that the curved surface of the conical taper results in a wave front distortion of the launched SPP field. The implications of such a distorted wave front for the field reaching the taper apex are illustrated in Fig. 2. Here we assume that we launch a 10-fs pulse onto the line grating and plot in Fig. 2(a) a snap shot of the resulting spatial SPP field distribution at a finite moment in time. The pulse is incident with a Gaussian intensity distribution and a flat wave front from $x$ direction. The real part of the electric-field amplitude is color-coded as a function of the distance from the apex and the angle $\varphi$, showing the distorted wave fronts as given by Eq. (3). The resulting SPP field near the taper apex is given as a coherent superposition of all SPP waves [19, 20]. Its time structure can be estimated by integrating the field at distance $r = c_{SPP}t$ from the taper apex over all angles $\varphi$ ($c_{SPP}$: SPP phase velocity) The result is shown in Fig. 2(b), where the temporal electric field amplitude of the distorted SPP pulse is plotted in black and with its pulse center at 30 fs, in comparison to a pulse with an ideal, flat wave front plotted in blue at $t = 0$ fs. Wave front distortion leads to destructive interference and hence to a reduced temporal electric field amplitude and, consequently, to a lower peak intensity (see Fig. 2(c)). The analysis given above implicitly assumes that the SPP field excited by the incident light propagates with the same characteristics as the lowest order, fully symmetric $n = 0$ eigenmode of the conical taper [2, 3, 21, 22]. Both, spatially inhomogeneous illumination of the grating and wave front distortion will enhance the coupling to higher order modes of the taper, which show different propagation characteristics: Such modes are not or only weakly guided towards the taper apex and hence their excitation will necessarily reduce the coupling efficiency [14]. The model outlined above also assumes an idealized taper geometry, and scattering losses due to a possible roughness of the taper surface are neglected.

*2.2 Back propagation of SPP waves from the tip apex*

The destructive interference of the SPP electric field near the taper apex can in principle be avoided by tailoring the spatial phase or wave front curvature of the incident light field. In order to estimate an optimized spatial phase profile, we consider the phase acquired by a SPP field that is propagating from the tip apex towards the grating at $\vec{r}_{inc} = (r_{inc}, \alpha, \varphi)$. Essentially, we assume a single dipolar emitter localized to a few nm at the tip apex, giving rise to a monochromatic SPP field propagating along the tip shaft in back direction, i.e., from the apex to the slit [19]. Then the phase of the SPP field is given by

$$\phi_{SPP}(\vec{r}) = \int_0^{\vec{r}} \vec{k}_{SPP}(\varphi) d\vec{r} = k_{SPP} \cdot r \tag{4}$$

In contrast to the incident light field, the SPP phase depends only on the distance $r$ from the apex. At any position $\vec{r}$ on the tip surface a phase mismatch $\Delta\phi(\vec{r})$ arises between the phase profile created by a plane incident wave (Eq. (2)) and that resulting from SPP back propagation. This amounts to

$$\Delta\phi(\vec{r}) = k_L \cdot \sin\alpha \cdot (r\cos\varphi - r_{inc}) - k_{SPP,r} \cdot (r - r_{inc}). \tag{5}$$

In an ideal setup, this phase or wave front mismatch should be compensated in order to maximize constructive interference at the apex and thus to attain the highest possible intensity at the nanofocus. In general, this can be achieved by optimizing the geometric shape of the grating coupler, by tailoring the spatial phase profile of the incident far field light or by combining both approaches.

We first consider the implications of Eq. (5) for an optimum coupler geometry. It suggests that the phase shift $\Delta\phi$ can be minimized by fabricating a grating with a curved shape, given by the distance $r_{gr}$ as a function of the angle $\varphi$ of

$$r_{gr}(\varphi) = r_{inc} \cdot \frac{k_{SPP,r} - k_L \cdot \sin\alpha}{k_{SPP,r} - k_L \cdot \sin\alpha \cdot \cos\varphi}. \tag{6}$$

This optimum grating shape is indicated as a black dashed line in Fig. 3. Even though the fabrication of such a designer grating seems in principle possible, several issues make this approach highly impractical: Both the opening angle and radius of curvature of the taper at the grating location need to be precisely known before the grating can be designed. Moreover, the tip opening angle usually is not constant over the entire propagation distance, and surface roughness scattering is expected to alter the SPP propagation in a way that is difficult to predict.

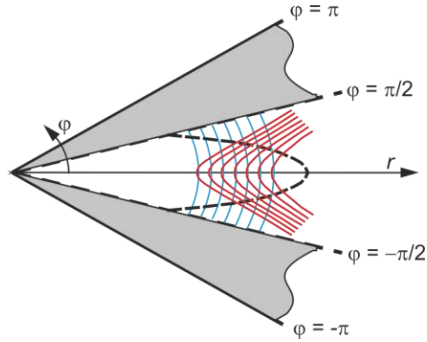

Fig.3. Unwrapped tip surface. SPP wave fronts resulting from grating coupling a plane wave incident field via an infinitely narrow single slit are shown as red curves, and the ideal wave fronts given by SPP back propagation are shown as blue curves. The mismatch could be corrected by a curved grating as indicated by the black dashed curve.

A second possibility to achieve phase matching is to continue writing straight grating grooves and to instead adapt the wave front of the incident laser field to the ideal SPP field. Eq. (4) suggests that ideally, the wave front of the incident light should be essentially flat along the taper axis (*z*), whereas a curved wave front, matching the radius of curvature of the taper should be chosen along the *y*-direction. This is illustrated in Fig. 4. Experimentally, a microscope objective (with numerical aperture $NA = 0.5$) is used to couple the far field light onto the grating. Taking a Gaussian-shaped incident beam propagating along the *x*-direction, the local radius of curvature of the wave front is given as $R_G(x) = x\left[1 + (x_R/x)^2\right]$. Here $x_R = \pi w_0^2 / \lambda$ denotes the Rayleigh length and $w_0 = 2c/(\omega \cdot NA)$ the beam waist of the Gaussian spot. As illustrated in Fig. 4, wave front matching may be achieved by using an astigmatic focal spot with the beam waist along the taper axis positioned at the taper surface. The beam waist along the *y*-axis, however, should be displaced towards the center of the taper such that $R_G(x) = r_{inc} \sin \alpha$. This then results in a clearly astigmatic beam profile at the taper surface, stretched along the *y*-direction.

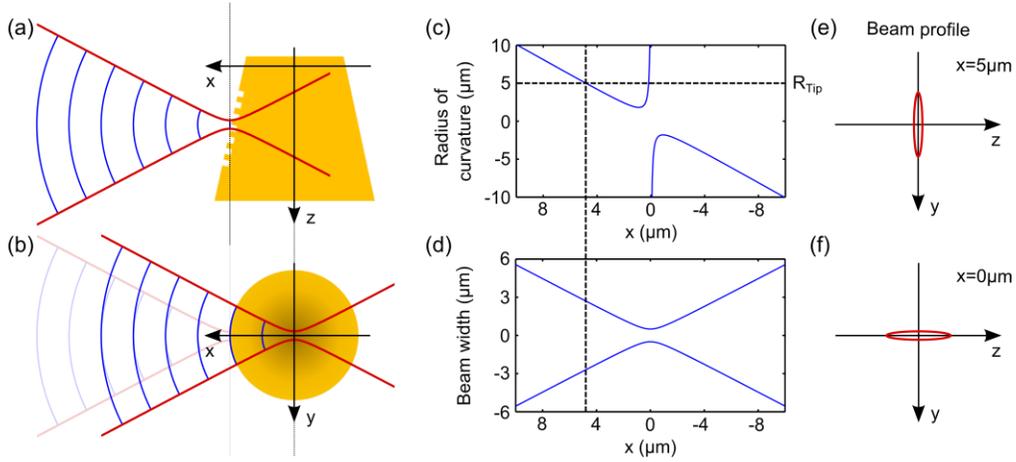

Fig. 4. Ideal wave front of the light field at the tip surface. To achieve phase matching, the wave front of the incident light should be (a) flat along the taper axis, *z*, and (b) curved along the *y*-direction. (c) Incident laser light wave front radius of curvature. From (c) and the tip radius as indicated by the broken black line results (d) the beam width at the tip surface. The resulting astigmatic focal spot with (e) a stretched beam profile along the *y*-direction at the taper surface and with (f) a stretched beam profile along the *z*-direction at the center of the taper.

The angular spectrum representation of a p-polarized Gaussian focal spot can be written as

$$\vec{E}_L(\vec{r}) = \iint dk_y dk_z \tilde{E}(k_y, k_z) e^{i\phi(k_y, k_z)} e^{i\vec{k}\vec{r}} \hat{e}(k_x, k_y) , \qquad (7)$$

where $\tilde{E}(k_y, k_z)$ denotes the amplitude and $\phi(k_y, k_z)$ the phase of the plane wave with wave vector $\vec{k} = (k_x, k_y, k_z)$, polarized along $\hat{e}(k_x, k_y)$. From this representation it can be seen that the desired astigmatic focus with a beam waist displaced along the *y*-direction can be generated by introducing a finite amount of spatial dispersion $\phi(k_y, k_z) = \frac{1}{2} \frac{\partial^2 \phi}{\partial k_y^2} \cdot k_y^2$ along the *y*-direction.

In the experiments described in the next section, we have therefore chosen to use a regular, straight grating shape while adapting the wave front of the incident light. For wave front shaping, a deformable mirror was used, and the surface profile of the mirror was optimized by an evolutionary algorithm. The advantages of this technique are a superior flexibility regarding the tip shape, tolerance with respect to slight misalignment, and, in addition, the possibility to correct for additional wave front distortions caused by optical elements in the beam path.

## 3. Experimental realization

*3.1 Experimental setup*

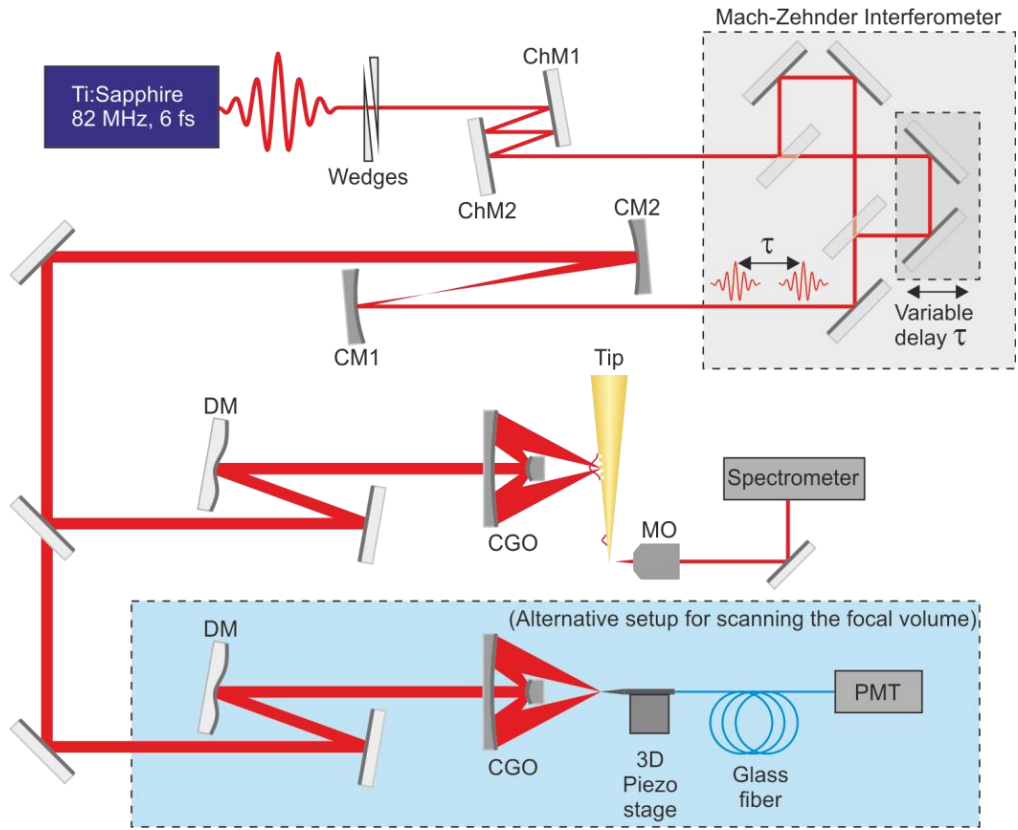

Fig. 5. Experimental setup. Ultrafast pulses from a 6-fs-Ti:Sapphire oscillator with chirp compensation (wedges and chirped mirrors ChM1 and ChM2) are split into two replicas by a Mach-Zehnder interferometer and the beam is expanded using a telescope (curved mirrors CM1 and CM2). After reflection off the deformable mirror (DM), the pulses are focused and grating-coupled onto the gold taper by a Cassegrain objective (CGO). The light scattered off the taper apex is collected with a microscope objective (MO) and a spectrometer. In the alternative setup (lower part, blue shaded box) the focus is scanned with a fiber probe on a three-axis piezo stage and a photomultiplier tube (PMT).

The experimental setup used in our work is schematically shown in Fig. 5. A commercial Ti:Sapphire oscillator (Femtolasers, Rainbow) generates few cycle laser pulses with a pulse duration of 6 fs at a repetition rate of 82 MHz and a pulse energy of 2.6 nJ. The oscillator is followed by a set of chirped mirrors (ChM1 and ChM2) with a group delay dispersion of -45 fs²/bounce (Femtolasers, GSM014) and a pair of wedges (Femtolasers, Suprasil, 2°48´) to compensate for dispersion of the laser mirrors and to pre-compensate for optical elements in the beam path. In order to measure the pulse duration, a pair of collinearly propagating pulses is generated through a dispersion-balanced Mach-Zehnder interferometer. The time delay between the two pulses is controlled by a single-axis piezo scanner (Physik Instrumente, P-621.1CD PI Hera). The beam diameter is extended by an all-reflective Kepler telescope (CM1 and CM2 in Fig. 5) to a beam size of 10 mm, and an aperture of 50 μm diameter is inserted at the beam waist for mode cleaning. An adjustable spatial phase shift is applied via a deformable mirror (DM, OKO-Tech, 15 mm 37-ch) with 10 mm usable aperture before focusing the pulses onto the gold tip by an all-reflective, aluminum-coated Cassegrain microscope objective (CGO, Davin Optronics, 5004-000) with an NA of 0.5, a 36x magnification and a working distance of 8.6 mm. The conical gold tapers were fabricated by electrochemical etching from single-crystalline gold wire. They have opening angles $\alpha$ of about 10-15° and tip radii of typically around 10 nm [9]. A nanoslit grating was ion-beam-milled onto the taper shaft for efficiently coupling far field laser light to SPPs [6]. We chose a grating period of 800 nm and slits with a width of 200 nm and a depth of about 100 nm. The light emitted from the localized plasmon forming at the apex is collected by a 0.7 NA glass microscope objective and the spectrum is measured using a spectrometer consisting of a 500-mm focal length monochromator (Princeton Instruments, Acton SP2500) and a nitrogen-cooled CCD detector (Princeton Instruments 100BR). This spectrometer was used to measure the fundamental spectra as well as the second harmonic (SH) generated at the tip apex. The SH radiation induced by illuminating the slit grating with the temporally delayed pulse pairs from the Mach-Zehnder interferometer enabled recording interferometric frequency-resolved autocorrelation (IFRAC) traces [23-25]. From these spectra, the time structure of the localized SPP electric field at the taper apex can be reconstructed.

To investigate the three-dimensional spatial extent of the focal volume, the gold tip and spectrometer were replaced by an aluminum-coated SNOM fiber probe (VEECO) (alternative beam path shown in the lower part of Fig. 5). The fiber probe was fabricated by focused ion beam milling to have an aperture diameter of ~300 nm and was mounted on a hardware linearized three-axis piezo stage (Physik Instrumente NanoCube) with a positioning accuracy of better than 10 nm. The fiber probe was scanned through the focus, and the locally collected light intensity was measured using a photomultiplier tube (PMT), similar to experiments presented in [26].

*3.2 Deformable mirror and evolutionary algorithm*

The DM consists of an aluminum-coated membrane mounted over a hexagonal array of 37 electrostatic actuators. In order to shape the wave front to the desired curvature, for each of the actuators the optimum position has to be found, which is a highly nonlinear problem due to the actuators' interdependence. For determination of the global optimum an EA is employed, and prior to that a suitable set of basis vectors is identified.

A priori, finding the global optimum of the mirror shape means finding one coordinate within a 37-dimensional parameter space, which makes the problem overly complex and slows down the procedure. Furthermore, the mirror is constructed as a single, continuous surface controlled with electrostatic actuators that allow only to pull and not to push the membrane. This excludes large portions of this parameter space, namely all shapes with a large derivative of the surface curvature.

A much more favorable approach than addressing each element individually is to use Zernike polynomials as basis vectors of a new parameter space. Zernike polynomials are prevalently used to describe wave-front curvatures in geometrical optics, e.g., from aberrations of optical elements [27]. Generally, a hexagonal structure as used here is not an ideal geometry to work with Zernike polynomials, and recent investigations have led to optimized keystone designs [28]. Here, however, we chose to use a commercially available deformable mirror with hexagonal design and to compromise on the purity of the generated patterns. In principle, Zernike polynomials represent a complete, orthonormal set of functions and can thus cover the identical parameter space as available when individually addressing the 37 mirror elements. However, much of the potential of the DM can already be exploited when truncating after the fourth order polynomials, and in practice it was found that truncating after the third order polynomials instead of the fourth already led to comparable results. Furthermore, there is no loss of performance when omitting the zero order, which corresponds to a mirror displacement, and even the first order, which corresponds to a tilt and can easily be compensated by the flat steering mirrors. Therefore, all experiments presented in the following have been obtained with a basic vector set consisting of the Zernike polynomials of order two and three, i.e., of seven polynomials. We describe a mirror curvature in this reduced, seven-dimensional parameter space by the corresponding $(7 \times 1)$ vector, where the vector components are the weighting factors for the seven considered Zernike polynomials, which are, in the single-index notation of Wyant and Creath [29]: Defocus ($Z_3$), Astigmatism 0° and 45° ($Z_4$ and $Z_5$), Coma X and Y ($Z_6$ and $Z_7$), and Trefoil 0° and 30° ($Z_9$ and $Z_{10}$).

The EA used to optimize these vector components is similar to the one introduced in Ref. [30]. The first generation for the EA is a population of twelve individuals, i.e., twelve vectors of random numbers in the range of $[-1,1]$. The four individuals giving the largest values of the fitness parameter survive and, together with four children created by mutation and four children created by cross-over, constitute the next generation. This cycle is typically repeated until the fitness saturates, which in our case starts before the tenth cycle, such that we terminated the EA after 30 cycles.

For mutation, a vector of random numbers is added to a surviving vector. Each of the random vector components is in the range $[-\beta, \beta]$, where $\beta = 0.2 \cdot e^{-n_g/10}$ decreases as a function of the generation number $n_g$. For cross-over, the four survivors are paired, and between each pair randomly chosen vector components are swapped, creating two pairs of complementary children. In a geometrical picture, such a vector pair spans a seven-dimensional cube. By cross-over, two diametral corners of this 128-cornered cube are selected as new vectors.

**4. Experimental results**

As the fitness parameter for the EA, we take the intensity radiated from the tip apex, which is obtained by integrating over the spectral power density measured by the spectrometer. Per iteration, the intensity is measured for each of the twelve individuals, and the four fittest are selected for the following cycle of the evolutionary algorithm. Figure 6(a) shows a typical evolution of the measured intensity maximum of the fittest individual of each generation (red squares). With each generation, a reference measurement is performed with a flat mirror surface (black squares). Note that what we term a flat mirror shape is the mirror with each actuator set to half the maximum voltage, and that the setup was adjusted manually with the DM in this defined state prior to optimization with the EA. The fitness, i.e., the integrated spectral power density of both the fittest individual and the reference in Fig. 6(a) are normalized to the average flat mirror fitness parameter. Already during the first generation, the fitness parameter increases by a factor of 3. During the optimization, the intensity increases quickly and converges already after a few generations. In the shown example, 90% of the final fitness is reached after the 7[th] generation, and in total the radiated intensity is enhanced by a factor of 8. The duration of the complete procedure is mainly determined by the exposure time of the spectrometer, which was set to 100 ms, and amounts to less than one minute.

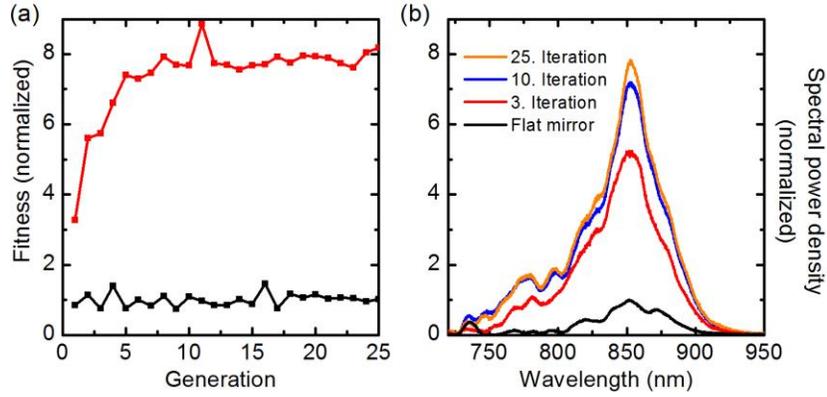

Fig. 6. (a) Fitness (integrated spectral power density) of the fittest individual of each generation (red squares) in comparison to the value measured with a flat mirror profile (black squares). (b) Recorded spectra of the fittest individual of the 3$^{rd}$, 10$^{th}$, and 25$^{th}$ generation in comparison to the spectrum recorded for a flat mirror (black curve).

Note that the curve displayed in Fig. 6(a) does not increase monotonously, as it is ideally expected from an EA. Slight fluctuations in intensity occur, which we ascribe to mechanical instabilities, mainly due to thermal expansion of the gold tip and due to drift of the three-axis piezo stage carrying the tip.

In Fig. 6(b) exemplary spectra are shown, which were recorded during the optimization procedure. The displayed spectra are those recorded for the fittest candidate of the 3$^{rd}$, 10$^{th}$, and 25$^{th}$ generation, respectively, in comparison to the spectrum recorded with a flat mirror (black curve in Fig. 6(b)), and normalized to the maximum value of this spectrum. From Fig. 6(b) it can be seen that the over-all spectral shape of the radiated light is well preserved during the optimization procedure, and the enhancement of the spectral maximum is in agreement with the 8-fold enhancement of the integrated value.

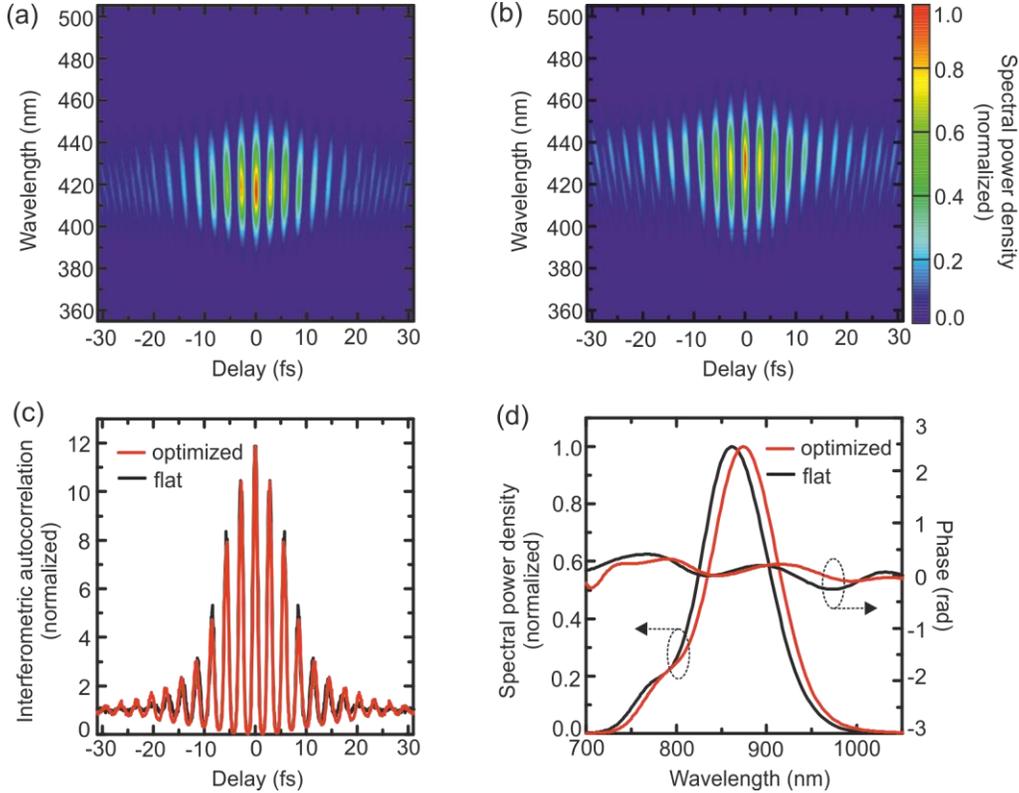

Fig. 7. Pulse duration. (a) IFRAC traces measured with a flat mirror and (b) with the optimized shape of the DM after the optimization procedure. (c) Interferometric autocorrelation traces extracted from the IFRAC traces obtained with the flat mirror (black curves) and with the optimized curvature (red curves), and (d) retrieved spectral intensity and phase.

The time structure of the electric field scattered from the taper apex was measured by IFRAC [23, 24], directly utilizing the SH generated at the tip apex [31]. Recorded IFRAC traces for a flat mirror and for an optimized mirror shape are shown in Figs. 7(a) and 7(b), respectively, where the SH spectrum in the range between 355 and 505 nm is displayed as a function of the time delay of the Mach-Zehnder interferometer. Interferometric autocorrelation (IAC) traces are extracted from the IFRAC traces by integration along the wavelength axis and are plotted as the black and the red curve in Fig. 7(c) for the flat and the optimally curved mirror, respectively. The full width at half maximum (FWHM) of both curves is about 12 fs. Note that the minimum:background:maximum ratio is 0:1:12 instead of 0:1:8 as is usual for SHG autocorrelations, which we ascribe to a contribution of nonlinear effects of higher order to the measured signal from the tip apex [32]. The retrieved spectral intensities and spectral phase profiles plotted in Fig. 7(d) are also similar. The slight wavelength offset of the two spectra of 13 nm is most probably due to a change in alignment: a shift of about 20 nm of the SPP resonance is expected when the angle of incidence of the incoming laser beam on the grating coupler is changed by about 1° [6]. This could be easily caused by the DM changing the surface curvature. For both spectra, however, the spectral phase curves are essentially flat. Even the small residual phase fluctuations are very similar and result in a pulse duration (intensity FWHM) of 10 fs in both cases. Together, the IFRAC measurements demonstrate that both the spectral and temporal shapes are well preserved during the optimization procedure.

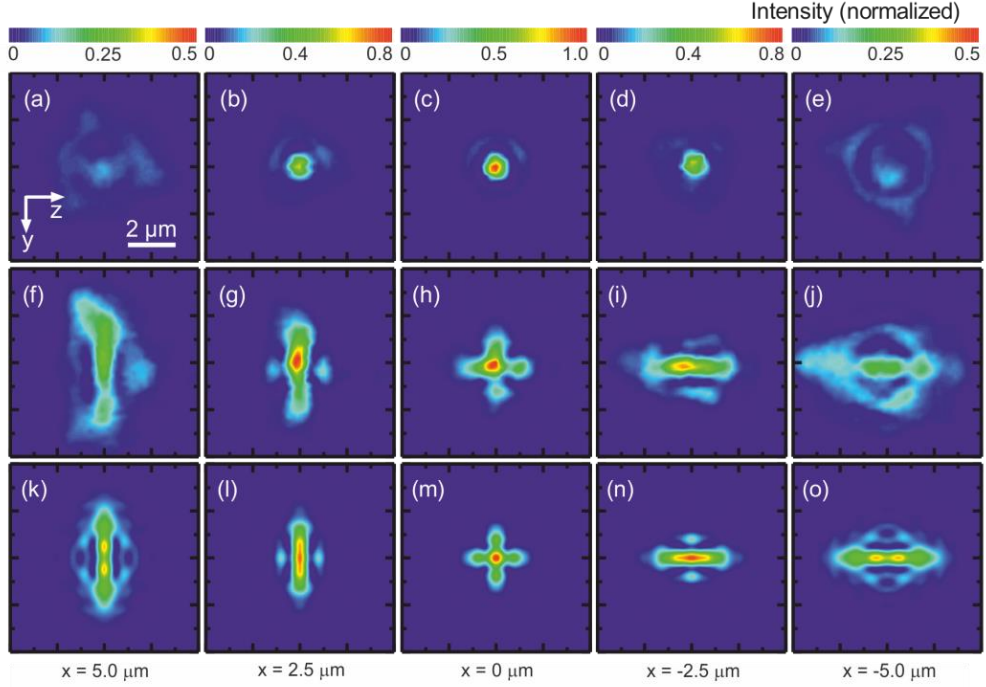

Fig. 8. Spatial intensity profiles of the incident laser light near the focus of the CGO. The profiles are measured by scanning a near field fiber probe with an aperture diameter of ~300 nm through the focus. Shown are cross sections along the *y-z* plane at five different positions along the *x*-axis. (a)-(e) Measured intensity distribution near the focus of the CGO when the DM was set to a flat front, (f)-(j) measured intensity distribution for the DM with the optimized front curvature, and (k)-(o) corresponding cuts through the focus with the calculated spatial phase profile.

Finally, we have measured the spatial intensity of the incident laser spot near the focus of the Cassegrain objective in order to characterize the wave front curvature created through the EA. The investigation of the focus volume was performed with the alternative setup in the lower part of Fig. 5 by scanning a near field fiber with an aperture of 300 nm through the focus. Some representative cuts through the recorded three-dimensional intensity maps are displayed in Fig. 8 for the DM with a flat front (Figs. 8(a)-8(e)) and for the optimized DM shape (Figs. 8(f)-8(j)). For a flat mirror, we see an approximately Gaussian-shaped focal shape, with its focal plane defined as $x=0$ (Fig. 8(c)). The diameter of the beam waist was measured to be <800 nm (intensity FWHM), in good agreement with expectations for a CGO with an NA of 0.5 [26]. When defocusing in forward ($x>0$) and backward ($x<0$) direction, we see a symmetric reduction in peak intensity, a concomitant increase in beam waist and a slight triangular shape of the intensity distribution, which is typical for the geometry of the CGO (see Figs. 8(a) and 8(e)).

For the deformable mirror settings which give the maximum light scattering from the tip apex, a clearly astigmatic beam shape emerges. The foci along the two axes ((g) and (i)) are displaced by approximately 5 µm along the *x*-axis. In ((f)-(j)), $x=0$ was defined as the position between the two foci. Experimentally we found that the optimum light scattering in Fig. 6 was reached with an intensity profile roughly corresponding to that in (g), i.e., when positioning the beam waist along the *z*-axis close to the taper surface, Hence, the wave fronts along the *y*-axis are bent towards the taper surface, as illustrated in Fig. 4. Simulations of the spatial intensity profile ((k)-(o)) based on the angular spectrum representation in Eq. (6) are in reasonable agreement with these measurements when introducing the appropriate amount of spatial dispersion along the y-axis, $\phi = 0.37 \cdot rad \cdot \mu m^2 \cdot k_y^2$. These simulations indicate clearly that optimum light scattering from the taper apex is reached when using an astigmatic focal spot with a focal plane along the *z*-axis lying on the tip surface, while the focus along *y* is displaced by 5 μm towards the center of the taper. This agrees well with the predictions of Fig. 4 since the diameter of the gold taper near the grating position is about 10 μm. A more detailed comparison of Figs. 8 ((f)-(j)) and ((k)-(o)) indicates that also some higher order phase distortions have been introduced during the course of the evolutionary optimization. This shows that small additional phase distortions have been corrected by the DM. Such distortions might result from a non-conical shape of the taper or from elastic SPP scattering during propagation along the taper, induced, e.g., by a finite amount of surface roughness.

**5. Summary and conclusions**

In summary, we have introduced an adaptive optics scheme to enhance the efficiency of adiabatic SPP nanofocusing on a conical gold taper. When focusing ultrashort light pulses onto a grating coupler on a conical gold taper, we find that the light scattering intensity from the taper apex can be greatly enhanced by adjusting the wave front of the incident light to the grating geometry. We demonstrate that for the chosen line grating, a highly astigmatic beam profile optimizes the nanofocusing efficiency, in good agreement with predictions from a simplified ray tracing model. This indicates that in the present experiments, wave front optimization mainly enhances the grating coupling efficiency. Apparently, the influence of elastic SPP scattering, resulting, e.g., from surface roughness on the taper, is of minor importance for the nanofocusing efficiency.

Our experimental results indicate that the introduced wave front optimization scheme based on a deformable mirror and an evolutionary algorithm is flexible, robust and fast, making it a highly versatile new tool for light coupling into nanometric spots in, e.g., metallic nanoresonators. For the coupling of spectrally narrowband light sources, the flexibility can in principle be further enhanced by replacing the deformable mirror with a spatial light modulator [33-35]. When working with spectrally broadband, ultrashort light pulses, however, the negligible spectral dispersion of the deformable mirror presents a crucial advantage as long as the effects of combined spatio-spectral dispersion on the resulting beam profile are weak. In the present example, the focusing of propagating SPP waves over mesoscopic distances, this approximation seems to be well justified. We therefore believe that the introduced wave front adaptation scheme is of considerable interest for enhancing the interaction of ultrafast, few-cycle pulses with metallic nanostructures, antennas and waveguides and is likely to find applications, e.g., in enhancing optical nonlinearities, EUV generation and electron emission.

**Acknowledgements**

Financial support of the work by the Deutsche Forschungsgemeinschaft (SPP1391 and DFG-NSF Materials World Network), the Korean National Research Foundation (Global Research Laboratory project, K20815000003, funded by the Korean Government) and the European Union (CRONOS) is gratefully acknowledged.